\documentclass[preprint,showpacs,showkeys,amsmath,amssymb,notitlepage]{revtex4}

\usepackage{dcolumn}
\usepackage{bm}
\usepackage{eurosym}
\usepackage{url}
\usepackage{array}
\usepackage{amssymb}
\usepackage{amsmath}

\begin{document}
\def\Z{\mathbb{Z}}
\def\R{\mathbb{R}}
\def\N{\mathbb{N}}
\def\Q{\mathbb{Q}}
\def\H{\mathbb{H}}
\def\C{\mathbb{C}}
\def\gpbar{_{\mathsf {g}}\hat{p}}
\title{Geometry of Financial Markets -- Towards Information Theory Model
of Markets}

\author{Edward W. Piotrowski}
 \email{ep@alpha.uwb.edu.pl}
  \homepage{http://alpha.uwb.edu.pl/ep/sj/index.shtml}
\affiliation{%
Institute of Mathematics, University of Bia\l ystok \\ Lipowa 41,
Pl 15424 Bia\l ystok, Poland
}%

\author{ Jan S\l adkowski}

 \email{sladk@us.edu.pl}
\affiliation{
Institute of Physics, University of Silesia \\
Uniwersytecka 4, Pl 40007 Katowice, Poland
}%

\date{\today}

\begin{abstract}
Most of parameters used to describe states and dynamics of
financial market depend on proportions of the appropriate
variables rather than on their actual values. Therefore,
projective geometry seems to be the correct language  to describe
the theater of financial activities. We suppose that the object of
interest  of agents, called here baskets, form a vector space over
the reals. A portfolio is defined as an equivalence class of
baskets containing assets in the same proportions. Therefore
portfolios form a projective space. Cross ratios, being invariants
of projective maps, form key structures in the proposed model.
Quotation with respect to an asset $\Xi$ (i.e.~in units of $\Xi$)
are given by linear maps. Among various types of metrics that have
financial interpretation, the min-max metrics on the space of
quotations can be introduced. This metrics has an interesting
interpretation in terms of rates of return. It can be generalized
so that to incorporate a new numerical parameter (called
temperature) that describes agent's lack of knowledge about the
state of the market. In a dual way, a metrics on the
 space of market quotation is  defined. In addition, one can define
 an interesting metric structure on the space of portfolios/quotation
 that is invariant with respect to
 hyperbolic (Lorentz) symmetries of the space of portfolios. The
introduced formalism opens new interesting and possibly fruitful
fields of research.
\end{abstract}

\maketitle

\section{\label{sec:level1} Introduction}
In majority  of the models considered in economics one cannot ask
questions about symmetries of the considered phenomena, especially
if one put the stress on group theoretical aspects. The reason is
that one can hardly speak about invariance (covariance) of terms
used in analysis or numerical values returned by most of models
\footnote{An invariant of a process (phenomenon, transformation
etc.) is a numerical parameter whose value remains constant during
that process. Analogously, a covariant is a parameter whose
numerical value has only a (relative) sense in a preselected
coordinate system but changes in a specified way  if the
coordinate systems is changed (e.g. given by some symmetry
transformation).}. We would like to argue that projective
geometry, equipped with an appropriate metric structure and some
measure of investors performance, might form a precise formalism
that allows us to carry out objective (quantitative) analysis of
investment processes and symmetries of their market context. We
describe a simple geometrical model of a financial market -- we
call it Information Theory Model of Markets (ITMM) -- that
explores ideas of projective geometry. Our model presents in some
sense a picture of financial markets  dual to that assumed in the
most popular ones, Capital Asset Pricing Model and Arbitrage
Pricing Models \footnote{D. G. Luendbeger, {\it Investment
Science}, Oxford University Press, New York 1998.}. Investors, due
to their lack of knowledge, wrong prognosis for the future or
simple fear, behave in an unpredictable, chaotic way. Prices are
determined by their decisions -- in the same way as the gas
pressure is determined by (chaotic) particles dynamics. A
non-random pricing of capital assets follows from investors
knowledge and possible random factors  cancel themselves due to
variety of strategies adopted by investors if the market is liquid
enough. The formalism of projective geometry allows us to carry
out analysis of invariant and covariant quantities. A detailed
axiomatic formulation of the model will be given elsewhere
\footnote{Most of the axioms takes the same form as in standard
models because they simple define market organization.}, here we
would like to present only some basic features. The paper is
organized as follows. In the next section we give some basic
definitions and describe mathematical tools we are going to use.
Then we show the importance of metric structures and give two
exemplary metrics. It follows that some important analogies with
physical theories can be expected. Finally,  we discuss a possible
connection between investors performance and knowledge about
markets measured by information theory means.

\section{\label{sec:leve2}Projective geometry as a formalism
describing investments}The market determines what goods are made
 and what products are bought and sold. We assume that objects of investors interest span a
$(N\negthinspace+\negthinspace1)$-dimensional vector space $G$
over the reals. Elements of this vector space are called {\em
baskets}. Let us fix some basis
$\{\mathsf{g}_0,\mathsf{g}_1,\dots,\mathsf{g}_N\negthinspace\}$ in $G$.
$\mathsf{g}_\mu\negthinspace\negthinspace\in\negthinspace G$, the
$\mu$-th element of the basis,
 is
called the $\mu$-th  {\em  asset}\/ (market good).  Assets,
although selected in an arbitrary way, are distinguished because
they are used for effective bookkeeping, accounting, market
analysis and so on. For any basket $p\negthinspace\in\negthinspace
G$ we have a unique representation
$$ p=\sum_{\mu=0}^N p_{\mu} \mathsf{g}_\mu.
$$ The coefficient
$p_\mu\negthinspace\negthinspace\in\negthinspace\mathbb{R}$ is
called the  $\mu$-th {\em coordinate of the basket}. A {\em
portfolio}\/ is defined as an equivalence class of non-empty
baskets (that is in $G\setminus \{0\}$) \footnote{We could also
take the empty basket into consideration but this would spoil the
projective space interpretation.}. Two baskets $p^\prime$ and
$p^{\prime\prime}$  are equivalent if and only if there exists
$\lambda\in{\mathbb R}$, such that
$$
\sum_{\mu=0}^N p^\prime_{\mu} \mathsf{g}_\mu = \sum_{\mu=0}^N
\lambda~ p^{\prime\prime}_{\mu} \mathsf{g}_\mu .
$$
Equivalently,
$$
(p^\prime_0,\dots,p^\prime_N)= (\lambda
p^{\prime\prime}_0,\dots,\lambda p^{\prime\prime}_N).
$$
If for a given portfolio  we have
$p_\mu\negthinspace\negthinspace\neq \negthinspace0$, then there
exists such a basket representing this portfolio that it contains
exactly a unit of  asset $\mathsf{g}_\mu$. Coordinates of this
basket,
$p\negthinspace=\negthinspace(p_0,\dots,p_{\mu-1},1,p_{\mu+1},\dots,p_{N}),
$ are called {\em inhomogeneous coordinates}\/ of the portfolio
$p$ with respect to  $\mu$-th asset. If
$p_\mu\negthinspace\negthinspace=\negthinspace0$,
$p\negthinspace=\negthinspace(p_0,\dots
,p_{\mu-1},0,p_{\mu+1},\dots ,p_N)$, then we say that  that the
portfolio   $p$ {\em is improper}\/ for the $\mu$-th asset. {\em
Market quotation}\/ $U$ in units of $\nu$-th asset is a linear map
$U(\mathsf{g}_\nu,~\cdot~)\negthinspace: G \rightarrow {\mathbb
R}$.
The map $U$ associates with a given portfolio $p$\, its current
value in units of $\mathsf{g}_\nu$:
\begin{equation}
\label{pipi-rzut} (U p)_\nu = U(\mathsf{g}_\nu,p)  =
\sum_{\mu=0}^N U(\mathsf{g}_\nu,\mathsf{g}_\mu) p_\mu,
\end{equation}
where $U(\mathsf{g}_\nu,\mathsf{g}_\mu)$ is the  price \/ of a
unit of  $\mu$-th asset given in units of  $\nu$-th asset.

\subsection{\label{sec:leve21}Basic definition and ideas}

We require that
$$ U(\mathsf{g}_\mu, U(\mathsf{g}_\nu, p)
\mathsf{g}_\nu) \mathsf{g}_\mu = U(\mathsf{g}_\mu, p)
\mathsf{g}_\mu $$
for $p$ and $\mathsf{g}_\mu$ and
$\mathsf{g}_\nu$ being exchangeable assets (that is
$U(\mathsf{g}_\mu,\mathsf{g}_\nu)\neq 0$ and
$U(\mathsf{g}_\mu,\mathsf{g}_\nu)\neq \pm\infty$, so inserting $p=
\mathsf{g}_\rho$ we get
\begin{equation}
\label{lacz-rzut} U(\mathsf{g}_\mu,\mathsf{g}_\nu)
U(\mathsf{g}_\nu,\mathsf{g}_\rho) =
U(\mathsf{g}_\mu,\mathsf{g}_\rho)
\end{equation}
for any $\mu$, $\nu$, $\rho$. Therefore quotations are transitive
\footnote{Note that this means that we treat all taxes, brokerages
etc\mbox{.} as liabilities and therefore as separate assets.
Models that have scale effects (projective symmetry is broken)
should have a dual description in terms of nontransitive
quotations.}. If we set
$\mu\negthinspace=\negthinspace\nu\negthinspace=
\negthinspace\rho$ in $(\ref{lacz-rzut})$ then we see that there
are two possibilities $
U(\mathsf{g}_\mu,\mathsf{g}_\mu)\negthinspace=\negthinspace1\;\;\text{or}
\;\;U(\mathsf{g}_\mu,\mathsf{g}_\mu)\negthinspace=\negthinspace0$.
The case
$U(\mathsf{g}_\mu,\mathsf{g}_\mu)\negthinspace=\negthinspace1$
implies projectivity of $U$: $ U((Up)_\mu\mathsf{g}_\mu)_\mu =
(Up)_\mu $. The case
$U(\mathsf{g}_\mu,\mathsf{g}_\mu)\negthinspace=\negthinspace0$
means that the  $\mu$-th asset is not subjected to quotation in
the market (one can only, for example, present somebody with such
an asset). For  $\mu\negthinspace=\negthinspace\rho$ we get
$U(\mathsf{g}_\mu,\mathsf{g}_\mu)\negthinspace=\negthinspace1$ and
therefore
 $$
U(\mathsf{g}_\mu,\mathsf{g}_\nu)=\bigl(U(\mathsf{g}_\nu,\mathsf{g}_\mu)
\bigr)^{-1}. $$
 In general, the quotation map can be represented
by a $(N\negthinspace+\negthinspace
1)\negthinspace\times\negthinspace
 (N\negthinspace+\negthinspace1)$  matrix with
$(\mu,\nu)$-th entry  given by
$U_{\mu\nu}:=U(\mathsf{g}_\mu,\mathsf{g}_\nu)$. The simplest way
of determining this matrix consist in selecting some asset that is
called {\em the currency}. Suppose that the asset $\mathsf{g}_0$
is selected as the currency. The matrix $U_{\mu\nu}$ is defined
uniquely by $N$ values  $u_k := U(\mathsf{g}_0,\mathsf{g}_k)$ for
$k\negthinspace=\negthinspace\negthinspace1,\dots,N$. (Note that
$U_{00}\negthinspace=\negthinspace1$). If $u_0:=1$, due to the
transitivity   $(\ref{lacz-rzut})$  all entries of $(U_{\mu\nu})$
are determined by the formula:
\begin{equation}
\label{hor-rzut} U_{\mu\nu} = u^{-1}_\mu u_\nu.
\end{equation}
Explicitly, we have Eq.~$(\ref{pipi-rzut})$
\begin{equation*}
 (U p)_\nu = \sum_{\mu=0}^N u_\mu p_\mu
u^{-1}_\nu.
\end{equation*}
For  $u_k\negthinspace\negthinspace=\negthinspace0$ Eq\mbox{.}
$(\ref{hor-rzut})$ remains valid if we set $u^{-1}_k:=0$.
Sometimes we have to rescale the prices $u_k$ in units
proportional to  $\mathsf{g}_0$, ( e.g\mbox{.} if $\mathsf{g}_0$
represents shares, after split, after currency denomination and so
on). Therefore it is convenient to identify  quotations $
U=(\lambda,\lambda\, u_1,\dots,\lambda\, u_N) $ for all
$\lambda\negthinspace\in\negthinspace{\mathbb
R}\negthinspace\setminus \negthinspace\{0\}$, that is introduce
homogeneous coordinates. We say that the portfolio $p$ {\em is
balanced } for the quotation $U$ if there is such an asset
$\mathsf{g}_\mu$, so that the value of $p$ in units of
$\mathsf{g}_\mu$ is  $0$, that is
$$
 (U p)_\mu = \sum_{\nu=0}^N
 U(\mathsf{g}_\mu,\mathsf{g}_\nu) p_\nu  =
 \sum_{\nu=0}^N
 u_\nu p_\nu u^{-1}_\mu  =  0.
$$
For quotation denominated in currency this formula simplifies to $
\sum_{\mu=0}^N u_\mu p_\mu = 0$. The linearity of these equations
allows for simple interpretations: portfolio $p$ is balanced if
the corresponding point belongs to the hyperplane representing
quotation $U$.
\begin{table}[h] {\sl
\begin{center}
\begin{tabular}{|r|c|c|}
\cline{2-3}
\multicolumn{1}{c|}{}&{\rm  MARKET}& {\rm  PROJECTIVE GEOMETRY}\\
\cline{2-3} \hline
$p$ & portfolio&  point\\
$U$ & quotation&  hyperplane\\
$U p=0$ & portfolio is balanced for $U$ & point $p$ lies in quotation hyperplane  \\
\hline
\end{tabular}
\end{center}}
\caption{Projective geometry dictionary}
\label{hrzut-tabi}
\end{table}
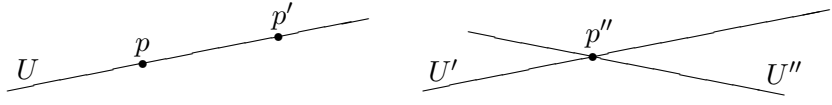
\begin{figure}[h]
\setlength{\unitlength}{.3cm}
\begin{picture}(18,3)
\put(0,0){\line(5,1){16}} \put(6.,1.2){\circle*{.3}}
\put(5.,1.5){\makebox(2,1)[b]{$p^{}$}}
\put(12.,2.4){\circle*{.3}}
\put(11.2,2.7){\makebox(2,1)[b]{$p^{\prime}$}}
\put(-0.1,0.5){\makebox(2,1)[b]{$U$}}
\end{picture}
\begin{picture}(15,3)
\put(0,0){\line(5,1){18}} \put(2,2.6){\line(5,-1){14}}
\put(7.5,1.5){\circle*{.3}} \put(6.8,1.9){\makebox(2,1)[b]{$p^{\prime\prime}$}}
\put(-0.1,0.4){\makebox(2,1)[b]{$U^\prime$}} \put(
15.0,.3){\makebox(2,1)[b]{$U^{\prime\prime}$}}
\end{picture}
\caption{ Two different portfolios $p$, $p'$ balanced for the same quotation $U$ and a portfolio $p''$ balanced for two different quotations $U'$, $U''$ (DUALITY!).}
\end{figure}

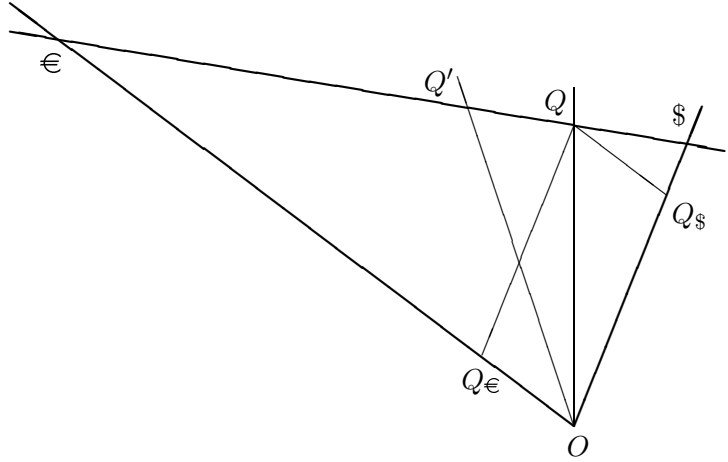
\begin{figure}[h]
\unitlength1mm
\begin{picture}(120,60)
\thicklines \put(80, 0){\line(-4, 3){75}} \put(80,
0){\line(2,5){17}} \put(80,40){\line(-6, 1){75}}
\put(80,40){\line(6,-1){20}} \thinlines \put(80, 0){\line( 0,
1){45}} \put(80, 0){\line( -1, 3){15.5}}
\put(80,40){\line(-2,-5){12.25}} \put(80,40){\line( 4,-3){12.3}}
\put(9,47){$\text{\euro}$} \put(93,40){$\$$} \put(76,42){$Q$}
\put(93,27){$Q_{\$}$} \put(79,-4){$O$}
\put(65,5){$Q_{\text{\euro}}$} \put(60,44.6){$Q'$}
\end{picture}
\caption{Exchange ratios.} \label{siankoo}
\end{figure}

An important invariant can be defined in projective geometry -- a
cross ratio of four points \footnote{H.~Busemann, P.~J.~Kelly.
{\em Projective Geometry and Projective Metrics}\/, Academic
Press, New York, 1953.}. For exchange ratios it describes the
relative change of quotation (cf Fig. $(\ref{siankoo})$:
$$
\{\$,Q,Q',\text{\euro}\}:=\frac{c'_{\$\rightarrow\text{\euro}}}{c_{\$\rightarrow\text{\euro}}}
=\frac{q'_\$ \, q_\text{\euro}}{q'_\text{\euro} \, q_\$}=
\frac{|Q'\text{\euro}|\,|Q\$|}{|Q'\$|\,|Q\text{\euro}|}=
\frac{P(\triangle_{Q'\text{\euro} O})\, P(\triangle_{Q\$ O})}{
    P(\triangle_{Q'\$O})\, P(\triangle_{Q\text{\euro} O})}\,,
$$
where $c_{\$\rightarrow\text{\euro}}:=\frac{q_\$}{q_\text{\euro}}$
is the exchange ratio ${\$\rightarrow\text{\euro}}$ (one obtains
for $q_\$ $ dollar $q_\text{\euro}$ euro) etc and
$P(\triangle_{abc})$ denotes the area of the triangle with
vertices $a,b$, and $c$. In Fig. $(\ref{siankoo})$ lengths of the
segments $Q\$$ and $Q\text{\euro}$ are proportional to $q_\$$ and
$q_\text{\euro}$, respectively. The invariance cross ratios of is
crucial to our model.
\subsection{\label{sec:leve22}Example: trading in a single asset}
Let us consider the cross ratio $[\mathfrak{G},U_{\to
\mathfrak{G}},U_{\mathfrak{G}\to},\$] $ for $ U_{\to
\mathfrak{G}}:=(v,v\, \mathrm{e}^{p_{\to
\mathfrak{G}}},\ldots)\;\;\text{and }\;\;
U_{\mathfrak{G}\to}:=(w,w\,
\mathrm{e}^{p_{\mathfrak{G}\to}},\ldots) $
 and the points
 $\mathfrak{G}$ and $\$$  given by crossing of the prime line
 $U_{\to \mathfrak{G}}U_{\mathfrak{G}\to}$ and one-asset portfolios:
 $\overline{\mathfrak{G}}$ i $\overline{\$}$ corresponding to
assets $\mathfrak{G}$ and $\$$. $ p_{\to \mathfrak{G}}$ and
$p_{\mathfrak{G}\to}$ are the logarithmic quotations for buying
and selling, respectively and (dots) $\ldots$ represent quotations
for the remaining assets and need not be the same for both
quotations. The logarithm of the cross ratio $[\mathfrak{G},U_{\to
\mathfrak{G}},U_{\mathfrak{G}\to},\$]$ on the straight line
$U_{\to \mathfrak{G}}U_{\mathfrak{G}\to}$ is equal to:
\begin{equation*}
\begin{split}
 \ln[\mathfrak{G},U_{\to
\mathfrak{G}},U_{\mathfrak{G}\to},\$]&= \ln[\frac{w\,
\mathrm{e}^{p_{\mathfrak{G}\to}}}{
         w\, \mathrm{e}^{p_{\mathfrak{G}\to}}-v\, \mathrm{e}^{p_{\to \mathfrak{G}}}},
         1,0,\frac{w}{w-v}]=\ln\frac{v\, w\, \mathrm{e}^{p_{\mathfrak{G}\to}}}{v\,
         \mathrm{e}^{p_{\to \mathfrak{G}}}\, w}= p_{\mathfrak{G}\to }-
         p_{\to \mathfrak{G}}.
\end{split}
\end{equation*}

\section{Metric structures}
It is a common lore that price movements are best described by
diffusion processes. Diffusion equations of various types involve
Laplace operator and therefore metric structure. Metric structures
are to some extent independent of the configuration (phase) space
structure. One of our aims is to find a suitable metrics on the
projective space. Various premises rooted in finance theory can be
used to select a metric structure on the space of portfolios. For
example, often we would like to know which market movements are
equivalent to portfolio modifications. Below we describe two
classes of metrics that we were able to construct in an explicit
way. Both have interesting physical connotations. There probably
is quite a lot of other interesting metrics yet to be found.
\subsection{Exemplary metric structure}
Let us try to define a metrics on the space of quotations. Two
different quotations $U^\prime$ and $U^{\prime\prime}$ determine
projective prime line. To define a cross ratio we need two
additional points lying in that line. It seems natural to select
them, let us consider two hyperplanes of improper quotations for
two basic assets. These hyperplanes cut the projective space
${\mathbb R\mathrm P}^N$ into $2^N$ $N$-dimensional simplexes.
Suppose that the quotations belong to the same simplex -- only
then the distance would be finite. Each hyperplane of improper
quotation for an asset $\mathsf{g}_\mu$ is cut by the prime line.
In this way we select $N\negthinspace+\negthinspace1$ points but
only two of them, say $P_b$ and $P_c$,  lie in the vicinity of
$U^\prime$ and $U^{\prime\prime}$ -- and only these two points
belong to the boundary of the simplex that contains $U^\prime$ and
$U^{\prime\prime}$, cf Fig. $\ref{hrzut-kuur}$. The cross ratio
$[P_b,U^{\prime},U^{\prime\prime},P_c]$ can be used to define the
distance (metrics):
\begin{equation}
\label{stopak-rzut} d(U^{\prime},U^{\prime\prime})
=\ln([P_b,U^{\prime},U^{\prime\prime},P_c]) = \ln
\frac{|U^{\prime}P_c||U^{\prime\prime}P_b|}
{|U^{\prime}P_b||U^{\prime\prime}P_c|}\,,
\end{equation}
 where $|P_1P_2|$ denotes euclidean distance of points $P_1$ and
$P_2$. After some tedious but elementary calculations the metrics
can be given in a more transparent form:
\begin{center}
 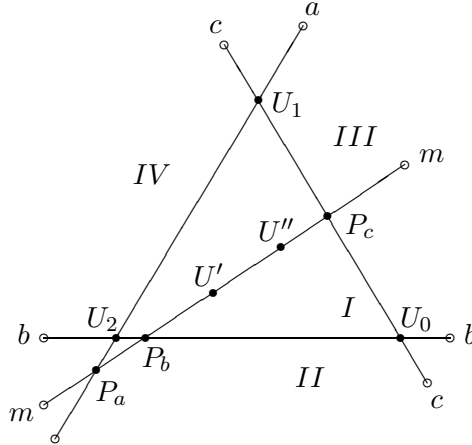
\begin{figure}[h]
 \begin{center}
\setlength{\unitlength}{.3cm}
\begin{picture}(20,18)
\put(-1.5,-2.5){\line(3,5){11}} \put(-1.5,-2.5){\circle{.3}}
\put(9.5,15.83){\circle{.3}}
\put(8.9,16.3){\makebox(2,1)[b]{$a$}} \put(15,0){\line(-3,5){9}}
\put(15,0){\circle{.3}} \put(14.4,-1.1){\makebox(2,1)[b]{$c$}}
\put(5.98,15.0){\circle{.3}} \put(4.6,15.5){\makebox(2,1)[b]{$c$}}
\put(-2,2){\line(1,0){18}} \put(-2.,2){\circle{.3}}
\put(-3.9,1.7){\makebox(2,1)[b]{$b$}} \put(16,2){\circle{.3}}
\put(15.9,1.7){\makebox(2,1)[b]{$b$}} \put(-2,-1){\line(3,2){16}}
\put(-2,-1){\circle{.3}} \put(-4.,-1.7){\makebox(2,1)[b]{$m$}}
\put(14,9.67){\circle{.3}} \put(14.2,9.8){\makebox(2,1)[b]{$m$}}
\put(1.2,2){\circle*{.3}} \put(-0.4,2.3){\makebox(2,1)[b]{$U_2$}}
\put(13.8,2){\circle*{.3}} \put(13.5,
2.3){\makebox(2,1)[b]{$U_0$}} \put(10.5,
2.9){\makebox(2,1)[b]{$I$}} \put(8.8,-.2){\makebox(2,1)[b]{$II$}}
\put(7.5,12.53){\circle*{.3}} \put(7.8,
11.8){\makebox(2,1)[b]{$U_1$}}
\put(10.8,10.5){\makebox(2,1)[b]{$III$}} \put(1.8,
8.8){\makebox(2,1)[b]{$IV$}} \put(2.5,2.){\circle*{.3}}
\put(2.,0.5){\makebox(2,1)[b]{$P_b$}} \put(.33,.56){\circle*{.3}}
\put(-.01,-1.){\makebox(2,1)[b]{$P_a$}}
\put(10.57,7.38){\circle*{.3}}
\put(11.0,6.4){\makebox(2,1)[b]{$P_c$}}
\put(5.5,4.0){\circle*{.3}}
\put(4.3,4.3){\makebox(2,1)[b]{$U^{\prime}$}}
\put(8.5,6.0){\circle*{.3}}
\put(7.3,6.3){\makebox(2,1)[b]{$U^{\prime\prime}$}}
\end{picture}
\end{center}
\caption{Quotation in a three-assets market ($\mathbb{R}P^2$).}
 \label{hrzut-kuur}
 \end{figure}
\end{center}
\begin{equation}
\label{minus-rzut}
\begin{split}
d(U^{\prime},U^{\prime\prime})& =\ln([P_b,U^{\prime},U^{\prime\prime},P_c]) = \ln
\frac{|U^{\prime}P_c||U^{\prime\prime}P_b|}
{|U^{\prime}P_b||U^{\prime\prime}P_c|}=
\ln\Bigl(\max_\mu\bigl(\frac{u^{\prime\prime}_\mu}
{u^{\prime}_\mu}\bigr)\Bigr) -
\ln\Bigl(\min_\mu\bigl(\frac{u^{\prime\prime}_\mu}
{u^{\prime}_\mu}\bigr)\Bigr)\\
&=\max_\mu\bigl(r_\mu(U^{\prime},U^{\prime\prime})\bigr) -
\min_\mu\bigl(r_\mu(U^{\prime},U^{\prime\prime})\bigr)= \max_\mu\bigl(r_\mu(U^{\prime},U^{\prime\prime})\bigr) +
\max_\mu\bigl(r_\mu(U^{\prime\prime},U^{\prime})\bigr).
\end{split}
\end{equation}
The function   $r_\mu(U^{\prime},U^{\prime\prime})$  is known in
finance  as
 {\em the interval interest rate.} We have already proposed a method that allows us to measure
 quantitatively
investors qualifications \footnote{E. W. Piotrowski, J. S\l
adkowski, Acta Phys.~Pol.~{\bf B32}  597 (2001).}. Inspired by
previous results and statistical physics, we can introduce a
temperature-like parameter in the metrics given by
Eq.$(\ref{minus-rzut})$.  Such a generalized metrics take the
following form:
\begin{equation*}
d(U^{\prime},U^{\prime\prime},T):=\frac{\sum\limits_\mu
r_\mu(U^{\prime},U^{\prime\prime})\,
\mathrm{e}^\frac{r_\mu(U^{\prime},U^{\prime\prime})}{T}}{\sum\limits_\mu
\mathrm{e}^\frac{r_\mu(U^{\prime},U^{\prime\prime})}{T}}+
\frac{\sum\limits_\mu r_\mu(U^{\prime\prime},U^{\prime})\,
\mathrm{e}^\frac{r_\mu(U^{\prime\prime},U^{\prime})}{T}}{\sum\limits_\mu
\mathrm{e}^\frac{r_\mu(U^{\prime\prime},U^{\prime})}{T}}\,.
\end{equation*}
It should be possible to define canonical ensembles of portfolios,
the temperature (entropy) of portfolios and, possibly, various
thermodynamics-like potentials in a way analogous to that of Ref.
[7].
\subsection{Hyperbolic (Lorentz) geometry}
We were able to identify another interesting metrics. Consider
quotations at two different times \( t^{\prime } \) and \(
t^{\prime \prime } \) in a simplified, two-assets market.  Let the
homogeneous coordinates are \( \hat{p}^{*\prime }=(\gpbar ^{\prime
}_{0},\gpbar ^{\prime }_{1}) \) and \( \hat{p}^{*\prime \prime
}=(\gpbar ^{\prime \prime }_{0},\gpbar ^{\prime \prime }_{1}) \),
respectively. Suppose the quotations are not equal,  \(
\hat{p}^{*\prime }\neq \hat{p}^{*\prime \prime } \). The linear
transformation:
 \[ \hat{S}=\hat{S}(\hat{p}^{*\prime },\hat{p}^{*\prime
\prime }):=\frac{1}{_{\mathsf {g}}\hat{p}^{\prime }_{0}\,
_{\mathsf {g}}\hat{p}^{\prime \prime }_{1}-\, _{\mathsf
{g}}\hat{p}^{\prime \prime }_{0}\, _{\mathsf {g}}\hat{p}^{\prime
}_{1}}\left(
\begin{array}{cc}
-\, _{\mathsf {g}}\hat{p}_{1}^{\prime }+\, _{\mathsf {g}}\hat{p}^{\prime \prime }_{1} & \, _{\mathsf {g}}\hat{p}^{\prime }_{0}-\, _{\mathsf {g}}\hat{p}^{\prime \prime }_{0}\\
-\, _{\mathsf {g}}\hat{p}^{\prime }_{1}-\, _{\mathsf
{g}}\hat{p}^{\prime \prime }_{1} & \, _{\mathsf
{g}}\hat{p}^{\prime }_{0}+\, _{\mathsf {g}}\hat{p}^{\prime \prime
}_{0}
\end{array}\right) \]
 changes the basis in such a way that the quotations  \(
\hat{p}^{*\prime } \) and \( \hat{p}^{*\prime \prime } \) have
coordinates   \( _{\mathsf {f}}\hat{p}^{\prime }:=(1,-1) \) and
 \( _{\mathsf {f}}\hat{p}^{\prime \prime }:=(1,1) \).
From the physicist point of view, the directions $(1,-1)$ and
$(1,1)$ define the propagation of light in a two-dimensional
spacetime. We can accept this directions us  absolute directions
(light cone). The underlying metric structure can also be found.
In the dual representation, that is in the space of portfolios,
two portfolios balanced on quotations $\hat{p}^{*\prime }$ and
$\hat{p}^{*\prime \prime }$ are infinitely separated. Explicit
form of the metrics on the space of portfolios is as follows:
\begin{equation*}
 d(p^{*\prime },p^{*\prime \prime
})=|\arctan (v^{\prime })-\arctan (v^{\prime \prime })|,
\end{equation*}
 where
\[
v(p^{*})=v(p^{*},\hat{p}^{*\prime },\hat{p}^{*\prime \prime
})=\frac{_{\mathsf {g}}p_{0}(_{\mathsf {g}}\hat{p}^{\prime \prime
}_{0}-\, _{\mathsf {g}}\hat{p}_{0}^{\prime })+\, _{\mathsf
{g}}p_{1}(_{\mathsf {g}}\hat{p}^{\prime \prime }_{1}-\, _{\mathsf
{g}}\hat{p}^{\prime }_{1})}{_{\mathsf {g}}p_{0}(_{\mathsf
{g}}\hat{p}_{0}^{\prime \prime }+\, _{\mathsf {g}}\hat{p}^{\prime
}_{0})+\, _{\mathsf {g}}p_{1}(_{ \mathsf {g}}\hat{p}_{1}^{\prime
\prime }+\, _{\mathsf {g}}\hat{p}^{\prime }_{1})}\,.
\]
Note that if we neglect details of the economic processes that
make capital then one can  change the content of a portfolio only
if one "travels with speed of light" in the market.
\subsection{Information theory context}
The projective geometry structure  of clear-cut market model with
a metrics that respects symmetries of the modelled processes
should yet be completed by discussion/construction  of
 algorithms that governs the supply and demand aspects of agents behaviour.
 These algorithms should be optimal from the metrical structure
 point of view  and, of course,  respect specific regulations laid down by
 authorities. For example, in the simple Merchandising Mathematician Model \footnote{E. W. Piotrowski, J. S\l
 adkowski,  Physica A 318 496 (2003).} and Kelly optimal bets \footnote{E. W. Piotrowski, M. Schroeder, Kelly Criterion revisited: optimal
 bets, talk given at the APFA5 Conference, Torino, 2006;
 physics/0607166.} the optimal market strategies  have direct
 connections with the Boltzmann/Shannon entropy. These examples
 suggest that there might be a unified description of market
 phenomena that involves tools from geometry, statistical physics
 and information theory. And the key ingredients would probably
 follow from the
 underlying metric structure.
\section{Conclusions: towards information theoretical description of markets}
We have attempted at formulation of kind of  Market Symmetry
Principle: Conclusions drawn from a logically complete market
model are invariant with respect to projective symmetry
transformations. We anticipate that metric structures might play a
key role that would pave the way for information theoretical
description of market phenomena. This point of view is supported
by the explicit examples given in the paper. The presented
projective geometry formalism although simplified, is, to the best
of our knowledge, the only one that attempts to introduce metric
structure to finance theory models that respect observed market
processes symmetries, eg  preselected absolute directions. This
would allow for analysis of hyperplanes of equilibrium
temperature, entropy, various thermodynamical potentials, Legendre
transforms and, possibly identification of conservation laws with
tools borrowed from information theory and (quantum) game theory
\footnote{E. W. Piotrowski, J. S\l adkowski, Int. J. Theor. Phys.
42 1101 (2003).}.



\end{document}